
\def\b{\beta}\def\d{\delta}\def\e{\epsilon}

\def\k{\kappa}\def\m{\mu}\def\n{\nu}\def\o{\omega}\def\t{\tau}

\def\O{\Omega}

\def\de{\partial}
\def\id{\equiv}\def\mo{{-1}}

\def\({\left(}\def\){\right)}\def\[{\left[}\def\]{\right]}

\def\pb{Poisson brackets }

\def\sch{Schwarzschild }
\def\poi{Poincar\'e }

\def\section#1{\bigskip\noindent{\bf#1}\smallskip}

\def\PL#1{Phys.\ Lett.\ {\bf#1}}

\def\PR#1{Phys.\ Rev.\ {\bf#1}}

\def\ref#1{\medskip\everypar={\hangindent 2\parindent}#1}
\def\beginref{\begingroup
\bigskip
\centerline{\bf References}
\nobreak\noindent}

{\nopagenumbers
\line{}
\vskip30pt
\centerline{\bf Cyclotron frequency and the quantum clock}

\vskip60pt
\centerline{
{\bf S. Mignemi}\footnote{$^\ddagger$}{e-mail: smignemi@unica.it}}

\vskip10pt

\smallskip
\centerline{$^1$Dipartimento di Matematica e Informatica, Universit\`a di Cagliari}
\centerline{viale Merello 92, 09123 Cagliari, Italy}
\smallskip
\centerline{$^2$INFN, Sezione di Cagliari, Cittadella Universitaria, 09042 Monserrato, Italy}

\vskip80pt

\centerline{\bf Abstract}
\medskip
{\noindent
We discuss the corrections to the orbital period of a particle in a constant magnetic field, driven by the
model of noncommutative geometry recently associated to a quantum clock.
The effects are extremely small, but in principle detectable.}
\vskip10pt
{\noindent

}
\vfill\eject\

}
\section{1. Introduction}
In the last years, models of noncommutative geometries have become one of the main areas of research in the field of
quantum gravity [1].
They can usually be obtained by a deformation of the commutation relations of standard quantum mechanics, yielding
a nontrivial commutator between spacetime coordinates.
The possibility that they could give rise to observable effects has also been largely studied [2].
Realistic experiments are based on astrophysical observations, but it is interesting to suggest different effects
that at least in principle could be observed in the laboratory.

Here, we describe a new effect, related to the motion of a particle in a constant magnetic field.
We shall consider a recently proposed model of noncommutative  geometry [3], inspired by the discussion of a quantum
clock [4], since in that case the calculations are particularly simple.
A quantum clock is a device that measures time through the decay of a sample of radioactive matter. Its nature entails
an uncertainty
relation between the interval of time measured and the size of the clock [4]. This uncertainty relation can be derived
in a relativistic quantum mechanical setting by postulating a deformation of the Heisenberg algebra, in which the time
and space coordinates have nontrivial commutation relations [3].

The main characteristic of this model is that only the time variable is deformed, which makes rather difficult to find
observable effects. The deformation can for example manifest itself only in time-related measurements in a relativistic
context.
A possibility of observing the deformation may however come from periodic motion. In [3] were for example considered
planetary orbits in a \sch background and corrections to the period of orbits predicted by general relativity were found.
However, it is unlikely that noncommutative geometries are effective in such macroscopic situations.

A more realistic experiment where such effects could be observable is the motion of a particle in a constant magnetic
field.
It is well known that in special relativity the orbits are circular, with frequency and radius depending on the
velocity (and hence the energy) of the particles.
As we shall show, in the quantum clock geometry the orbits are still circular, but their radius and frequency are
modified with respect to special relativity. The corrections for realistic experiments are extremely small, but in
principle might be detectable observing a great number of orbits.

The same effect can also occur in different models of noncommutative geometries, like Snyder [5] or $\k$-Poincar\'e [6],
but in those cases the calculations are much more involved.

\section{2. The model}
Let us review the geometry of the quantum clock [4]. For simplicity, we consider here its classical limit, where
commutators are replaced by Poisson brackets.
We denote $x^\m=(t,x_i)$, $p_\m=(E,p_i)$, with $i=1,2,3$ and signature $(-1,1,1,1)$.
The Poisson brackets read [4]
$$\eqalignno{&\{x_i,x_j\}=0,\qquad\{t,x_i\}=-\b{x_i\over r},\qquad\{p_\m,p_\n\}=0,\qquad\{x_i,p_j\}=\d_{ij},&\cr
&\{x_i,E\}=0,\qquad\{t,E\}=1,\qquad\{t,p_i\}={\b\over r}\(p_i-{x_jp_j\over r^2}x_i\).&(1)}$$
with $r=\sqrt{x_i^2}$ and, in the classical limit, $\b={G\over c^4}\sim 8\cdot 10^{-44}\ {\rm {sec^2\over m\,kg}}$,
with $G$ the gravitational constant. In the following we set $c=1$.

We study the relativistic motion of a charged particle in a static magnetic field.
Since the \poi algebra is undeformed in the quantum clock geometry [3], we take as Hamiltonian the standard
relativistic one,
$$H={1\over2m}(p_\m-eA_\m)^2={m\over2}\eqno(2)$$
and consider a time-independent vector potential, $A_\m=A_\m(x_i)$, with $A_0=0$.
Taking into account the \pb (1), the Hamilton equations read
$$m\dot x_i=p_i-eA_i,\qquad\qquad\dot p_i={e\over m}{\de A_j\over\de x_i}(p_j-mA_j),\eqno(3)$$
$$m\dot t=E+{\b\over r}\[\(p_i-{x_jp_j\over r^2}x_i-ex_j{\de A_i\over\de x_j}\)(p_i-eA_i)\],
\qquad\qquad\dot E=0,\eqno(4)$$
where a dot denotes a derivative with respect to the evolution parameter $\t$.
The only equation that differs from special relativity is the one for the derivative of the coordinate time.
It follows that $E$ is constant and can be identified with the energy of the particle.
Moreover,  eqs.~(3) give $\dot p_i=e{\de A_j\over\de x_i}\dot x^j$, and then
$$\dot u_i={e\over m}F_{ij}u_j,\eqno(5)$$
where we have defined $u_i=\dot x_i$.

Let us now consider a constant magnetic field in the $z$ direction, namely
$$F_{ab}=B\e_{ab},\qquad A_a=-{B\over2}\e_{ab}x_b,\qquad\qquad a,b=1,2.\eqno(6)$$
Then (5) becomes
$$\dot u_a={eB\over m}\e_{ab}u_b,\qquad\dot u_3=0.\eqno(7)$$
It follows that $u^2\id u_i^2=$ const.
With suitable initial conditions, the solution of (7) is given by
$$u_1=u\cos\o_0\t,\qquad u_2=u\sin\o_0\t,\qquad u_3=0,\eqno(8)$$
where $\o_0={eB\over m}$ and $u$ is an integration constant which coincides with the norm of $u_i$.
After integration, choosing the origin in the center of the orbit,
$$x_1={u\over\o_0}\sin\o_0\t,\qquad x_2=-{u\over\o_0}\cos\o_0\t,\qquad x_3=0.\eqno(9)$$
It follows that also $r$ is constant and equals ${u\over\o_0}$.
Till now, the solutions for the orbits are identical to those of special relativity. However, the relation between the parameter $\t$
and the time coordinate $t$ is now deformed. In fact, from (4) we obtain, using the solutions (8) and (9),
$${dt\over d\t}={E\over m}+{\b\over r}\(p_i-{x_jp_j\over r^2}x_i-ex_j{\de A_i\over\de x_j}\)\dot x_i={E\over m}+{\b\over 2}m\o_0u.\eqno(10)$$
Defining a constant $\O={m^2\o_0u\over E}={meBu\over E}$, we can write
$$t={E\over m}(1+\b\O)\t,\eqno(11)$$
and then
$$u_1=u\cos\o t,\qquad u_2=u\sin\o t,\qquad u_3=0,\eqno(12)$$
with
$$\o={m\o_0\over E(1+\b\O)}={eB\over E(1+\b\O)}.\eqno(13)$$
Moreover, we can now define the physical velocity as
$$v_i={dx_i\over dt}={mu_i\over E(1+\b\O)},\eqno(14)$$
which generalizes to our case the relativistic relation $v_i=mu_i/E$.

Starting from (14), we can write $\O$ as a function of $v=\sqrt{v_i^2}$. In fact,
$$u=(1+\b\O){Ev\over m},\eqno(15)$$
and substituting in the definition of $\O$, we obtain $\O=eB(1+\b\O)v$,
which can be solved for $\O$, giving
$$\O={\b eBv\over1-\b eBv}.\eqno(16)$$

At this point, it is useful to obtain the relation between the energy and the velocity of the particle.
To this end, we write the dispersion relation $E^2=m^2+p^2$ as
$$E^2=m^2(1+u^2)=m^2+E^2(1+\b\O)^2v^2,\eqno(17)$$
from which it follows
$$E^2={m^2\over1-(1+\b\O)^2v^2}={m^2\(1-\b eBv\)^2\over\(1-\b eBv\)^2-v^2}.\eqno(18)$$
Substituting in (13), we can finally write
$$\o={eB\over m}\ \sqrt{\(1-\b eBv\)^2-v^2},\qquad r={v\over\o}={mv\over eB\sqrt{\(1-\b eBv\)^2-v^2}}.\eqno(19)$$

These equations give the relation of the cyclotron frequency and orbital radius with the velocity of the particle.
They contain corrections of order $\b eB$ to the special relativity formulas. Expanding to first order in $\b$, one has for example
$$\o=\o_{\rm rel}\left(1-{\b eBv\over1-v^2}\right),\eqno(20)$$
where $\o_{\rm rel}={eB\over m}\ \sqrt{1-v^2}$ is the relativistic frequency.
From (17) one sees that the behavior of the frequency is similar to the one occurring in special relativity, but vanishes for a
velocity smaller than the speed of light, namely $v=(1+\b eB)^\mo$. The correction does not depend on the mass of the particle.

Eq.~(17) can also be inverted to obtain the velocity as a function of the energy,
$$v={\sqrt{E^2-m^2}\over E+\b eB\sqrt{E^2-m^2}},\eqno(21)$$
and substituting in (19), one can write the cyclotron frequency as a function of the energy of the particle.
The result is
$$\o={eB\over E+\b eB\sqrt{E^2-m^2}}\eqno(22)$$

As an illustrative example, one may consider a unit charge particle in a magnetic field of 8 Tesla with velocity $v\sim 0.99999999\,c$,
as typical of LHC.
In this case, the ratio $\o/\o_0$ is  of order $10^{-46}$. The predicted effect is extremely small, but one might exploit the cumulative
effect over several orbits to increase the efficiency of the measurement.

Although it is unlikely that corrections of this size can be detected, they are important from a theoretical point of view.
Similar effects can also be obtained in other noncommutative models, but the calculations are much more complicated in those cases.

\beginref
\ref [1] For a review, see P. Aschieri, M, Dimitrijevi\'c, P. Kulish, F. Lizzi and J. Wess, "Noncommutative spacetimes", Springer, Berlin 2009.
\ref [2] See for example G. Amelino-Camelia,   Symmetry {\bf 2}, 230 (2010).
\ref [3] S. Mignemi and N. Uras, \PL{A383}, 585 (2019).
\ref [4] L. Burderi, T. Di Salvo and R. Iaria, \PR{D93}, 064017 (2016).
\ref [5] H.S. Snyder, \PR{71}, 38 (1947).
\ref [6] J. Lukierski, H. Ruegg, A. Novicki and V.N. Tolstoi, \PL{B264}, 331 (1991).
\end